# Non-linear charge oscillation driven by single-cycle light-field in an organic superconductor

Y. Kawakami[1], T. Amano[1], Y. Yoneyama[1], Y. Akamine[1], H. Itoh[1],

G. Kawaguchi[2], H. M. Yamamoto[2], H. Kishida[3], K. Itoh[4], T. Sasaki[4],

S. Ishihara[1], Y. Tanaka[5], K. Yonemitsu[5], and S. Iwai[1*]

[1]Department of Physics, Tohoku University, Sendai 980-8578, Japan

[2]Institute for Molecular Science, Okazaki 444-8585, Japan

[3] Department of Applied Physics, Nagoya University, Nagoya 464-8603 Japan

[4]Institute for Materials Research, Tohoku University, Sendai 980-8577, Japan

[5]Department of Physics, Chuo University, Tokyo 112-8551, Japan

 71.27.+a, 74.25.Gz, 78.47.J-

*s-iwai@tohoku.ac.jp

Abstract

Intense light-field application to solids produces enormous/ultrafast non-linear phenomena such as high-harmonic generations [1,2] and attosecond charge dynamics [3,4]. They are distinct from conventional photonics. However, main targets have been limited to insulators and semiconductors, although theoretical approaches have been made also for correlated metals and superconductors [5]. Here, in a layered organic superconductor, a non-linear charge oscillation driven by a nearly single-cycle strong electric field of >10 megavolts /cm is observed as a stimulated emission. The charge oscillation is different from a linear response and ascribed to a polar charge oscillation with a period of ~6 fs. This non-linear polar charge oscillation is enhanced by critical fluctuations near a superconducting transition temperature and a critical end point of first order Mott transitions. Its observation on an ultrafast timescale of ~10 fs clarifies that the Coulomb repulsion plays an essential role in superconductivity of organic superconductors.

Light-induced charge motions in solids are governed by electron-electron scatterings; thus, a quasi-particle state by the dynamical screening effect [6-8]. These dynamics have been discussed in terms of an increase in the electron temperature in metals [9] and non-equilibrium quasi-particles in superconductors [10-14]. Recent advances in intense light-fields take us beyond such conventional descriptions in metals [15, 16] and superconductors [17, 18]. This makes us to expect a characteristic non-linear charge motion on a femtosecond-attosecond timescale reflecting a microscopic nature. They have been already reported in insulators [2, 4] and semiconductors [1, 3].

Since the discovery of high-transition-temperature (high-$T_{SC}$) superconducting cuprates, a phase diagram of a Mott transition, which consists of anti-ferromagnetic (AF)/Mott insulator, paramagnetic metal, and superconductor, has attracted much attention [19]. Criticalities near the Mott transition and superconductivity are promising candidates for exploring new optical non-linearities. κ-(BEDT-TTF)$_2$X (X; anion molecule) [20-23] is a layered organic superconductor with $T_{SC}$=11.6 K (Fig. 1a). The triangular lattice consisting of BEDT-TTF dimers is regarded as a 1/2-filled system, although the averaged charge per BEDT-TTF molecule is +0.5 (3/4 filling). The temperature-$t/U_{dimer}$ phase diagram [20-23] (Supplementary-1), where $t/U_{dimer}$ is the ratio of an inter-dimer transfer integral $t$ (proportional to the bandwidth) to the effective on-site Coulomb energy for a dimer $U_{dimer}$, is characterized by the first order Mott transition line and the critical end point ($T_{END}$~33 K, Supplementary-1) as shown in Fig. 1a. Superconducting fluctuations toward $T_{SC}$ below ~2$T_{SC}$ have been discussed in terms of a pseudogap [10-14].

In this letter, we report that a non-linear charge oscillation is induced in all of three κ-(BEDT-TTF)$_2$Cu[N(CN)$_2$]Br (*h*-Br, crystal and thin film) and κ-(*d*-BEDT-TTF)$_2$Cu[N(CN)$_2$]Br (*d*-Br, crystal) samples that were investigated (metallic to superconducting vs. insulating) by a 6 fs nearly single-cycle light-field of 11 MV/cm. The non-linear oscillation emerges as a stimulated emission in the near infrared region and is characterized as a polar charge oscillation (Fig. 1b). A distinctive enhancement near $T_{SC}$ and $T_{END}$ is shown only for the metallic-to-superconducting sample (*h*-Br crystal).

The optical conductivity spectrum (σ) of κ-(BEDT-TTF)$_2$Cu[N(CN)$_2$]Br (*h*-Br) along the *c*-axis at 4 K [24, 25] in Fig. 2a is well reproduced using the Drude-Lorentz model (orange line) [25, 26]. A large increase in the reflectivity R (ΔR/R~180% at 0.63 eV) is observed on the higher energy side of the dimer band for time delays ($t_d$) of 10 fs after the application of the light-field (along the *c*-axis, excitation intensity $I_{ex}$=1.0 mJ/cm$^2$ = 11 MV/cm)(Fig. 2b)(Supplementary-2). The rise in the electron temperature is negligible at $t_d$=10 fs because electrons are scattered only a few times before(Supplementary-3). The ΔR/R spectrum at $t_d$=200 fs in Fig. 2b is reproduced by the Lorentz analysis assuming an additional oscillator (see Fig. 2b's caption and Supplementary-2). To judge whether the observed ΔR/R peak at 0.63 eV is due to a stimulated emission or an induced absorption, we performed transient transmittance (ΔT/T, T denotes a transmittance) and ΔR/R measurements using a 180 nm thin film of *h*-Br (insulator, Supplementary-4) on a CaF$_2$ substrate [27] (Fig. 2c). The blue dots in Fig. 2c represent ΔT/T of the thin film (excitation and detection polarizations are along the $c$-axis, $I_{ex}$=1.0 mJ/cm$^2$, $t_d$=10 fs), which show a large increase (ΔT/T~20%) at 0.67 eV, $t_d$=10 fs. The ΔR/R spectrum (red dots in Fig. 2c) shows a reflectivity increase which is roughly consistent with that for the crystal (Fig. 2b). Note that ΔR/R>0 and ΔT/T>0 in Fig. 2c (optical gain) are ascribed to a stimulated emission (S. E.). The corresponding peak in ΔR/R is also observed for insulating $d$-Br (crystal) (Supplementary-4).

In the ΔR/R spectra for $t_d$=10 fs at 6 K (Fig. 3a), the peak is observed at 0.63 eV for any $I_{ex}$ (=1, 0.1 and 0.01 mJ/cm$^2$). The time profiles of the ΔR/R peak (Fig. 3b) at $I_{ex}$=1.0 mJ/cm$^2$ (red line) and at 0.1 mJ/cm$^2$ (blue line) are analysed assuming fast (< 10 fs) and slow (~100 fs) build-up components. They are reproduced by the conventional method as shown by the black line that has an oscillating component manifested in Fig. 3c. The details are described in Supplementary-5. The fast build-up (< 5 fs) with decay (70 fs) is dominant (65.4 %) in comparison with the slow rise (90 fs). However, the slow component is dominant (67.8 %) at $I_{ex}$=0.1 mJ/cm$^2$. The ultrafast component is strongly non-linear (Fig. 3d), i.e., ΔR/R at 0.63 eV for $t_d$=10 fs (red circles) shows a super-linear dependence on $I_{ex}$ (the dashed line shows an eye guide for a linear increase). For larger $I_{ex}$ in the inset, it continues to increase linearly up to $I_{ex}$>1.3 mJ/cm$^2$. The solid line intercepts at $I_{ex}$~0.1 mJ/cm$^2$, showing the threshold-like dependence for the ultrafast component.

The highlight is an enhancement of ΔR/R near $T_{SC}$ and $T_{END}$. The temperature dependence of the ΔR/R spectrum is shown for $t_d$=10 fs at $I_{ex}$ =1.0 mJ/cm$^2$ (Fig. 4a) and 0.01 mJ/cm$^2$ (Fig. 4b). The 0.63 eV peak grows with reducing temperature below 50 K for both $I_{ex}$. These ultrafast responses indicate an anomalous enhancement near $T_{END}$ (~30 K) and further increases at lower temperatures (1.0 mJ/cm², Fig. 4c) (Results for $t_d$=500 fs are in Supplementary-6). The rise in the electron temperature is negligible at $t_d$=10 fs, although the anomalies at ~10 K and ~28 K show 2~4 K shifts (at 1.0 mJ/cm²) to the low-temperature side because of a heat accumulation in a 1 kHz operation (Supplementary-3). This ~10 fs response is in contrast to the energy/time scale of a superconducting gap and its fluctuation (~meV from $T_{SC}$=11.6 K, ~picosecond). For small $I_{ex}$, the increase in ΔR/R below ~2$T_{SC}$ at $t_d$=10 fs becomes obscure (0.1 mJ/cm², Fig. 4d) and turns to decrease (0.01 mJ/cm², Fig. 4e), while the anomaly at ~$T_{END}$ remains. No anomalies are observed in the temperature dependence of R in *h*-Br (Supplementary-7) and that of ΔR/R in insulating κ-(*d*-BEDT-TTF)$_2$Cu[N(CN)$_2$]Br, which does not have $T_{SC}$ or $T_{END}$ on its small $t/U_{dimer}$ side (Supplementary-4).

To clarify the origin of the non-linear charge motion, we theoretically investigated charge oscillations driven by the single-cycle light-field in a two-dimensional (2D) 3/4-filled extended Hubbard model for a 16-site system (Fig. 5a). The details are in Supplementary-8. Fourier transform (FT) spectra of the charge-density time profile of a molecule are shown by the lower panel of Fig. 5c. The weight of the FT spectrum for the field amplitude $F$=0.01 [V/angstrom] along the *c*-axis, on-site Coulomb repulsion $U$=0.8 eV, nearest-neighbour Coulomb repulsions between sites *i* and *j* $V_{ij}$=0 (green dots in Fig. 5c) is mainly distributed (≤ 0.7 eV) in the spectral range of σ for polarization along the *c*-axis (upper panel of Fig. 5c). This FT spectrum for $F$=0.01 corresponds to charge oscillations due to transitions from bonding to anti-bonding states around a dimer. With increasing $F$, the spectral weight at a higher energy (~0.9 eV) becomes dominant. This FT spectral weight is attributable to the polar charge oscillation in Fig. 5b. This polar charge oscillation is driven by the charge transfers through different bonds, i.e., charges flow into a white site from four neighbouring gray sites through the $b_1$, $b_2$, and two $q$ bonds (Fig. 5b, left), and then charges flow back to the gray sites (Fig. 5b, right). The observed peak energy of 0.63 eV indicates that the oscillation period is ~6 fs.

At ~$T_{END}$, the potential barrier is absent for the phase transition. In such a situation, a strong light field can induce the non-linear charge response shown in Figs. 5a and 5b with large amplitude. Since the Coulomb repulsion is larger than 0.4 eV($=\hbar/(10fs)$), the ultrafast ~10 fs response of the anomaly at ~$T_{END}$ is reasonable, i.e., the insulator to metal transition is of electronic origin because the timescale of inter-molecular motion (> 200 fs) is much longer.

It is noteworthy that the instantaneous charge distributions (Fig. 5b) are consistent with a hidden polar charge order near the superconductivity in theoretical calculations by the random-phase approximation [28] and by a variational Monte Carlo method [29]. Essence of inducing such instantaneous charge distributions (similar to the hidden charge order) is field-induced simultaneous charge transfers through the inter- and intra- dimer bonds (Fig. 5b).

Though a detailed mechanism for the enhancement near $T_{SC}$ is unclear, the 10 fs response is sensitive to $T_{SC}$ only for the strong field (Fig. 4). This result indicates that the non-linear polar charge oscillation is promoted by a microscopic mechanism of the superconductivity with the energy scale > 0.4 eV from Coulomb repulsion. Thus, a relation between Coulomb repulsion and the non-linear charge oscillation is important for understanding the enhancement of ΔR/R near $T_{SC}$. Fig. 5d shows that the non-linear charge oscillation becomes more dominant with increasing $U$, indicating that Coulomb repulsion is indispensable to synchronize charge motions through the intra- and inter- dimer bonds and is essential for the superconductivity as well. The ultrafast response of the non-linear charge motion, which is much faster than a characteristic timescale of the superconducting gap (~ps), leads us to expect a non-perturbative control of many-body charge motion.

In summary, this report demonstrates the stimulated emission driven by the non-linear polar charge oscillation with a period of ~6 fs in the organic superconductor κ-(BEDT-TTF)$_2$Cu[N(CN)$_2$]Br. The stimulated emission is enhanced near $T_{SC}$ as well as $T_{END}$. A 10 fs response of this enhancement near $T_{SC}$ shows the essential role of Coulomb repulsion.

## Methods

**Sample preparation.** Single crystals of κ-(BEDT-TTF)$_2$Cu[N(CN)$_2$]Br (0.7 × 0.5 x 0.8 mm for axes *a, b, c,*), κ-(*d*-BEDT-TTF)$_2$Cu[N(CN)$_2$]Br (0.6 × 0.5 × 0.7 mm, *d*-BEDT-TTF denotes deuterated BEDT-TTF molecule) and a thin film of κ-(BEDT-TTF)$_2$Cu[N(CN)$_2$]Br (0.5 × 1.8×10$^{-4}$ ×0.5 mm on a CaF$_2$ substrate with the thickness of 0.5 mm) were prepared using the methods described in previous studies [24, 27].

**6 fs infrared pulse generation.** The 6 fs pulse covering 1.2–2.3 μm, is generated by the method described in Refs. 15 and 30, i. e., a broadband infrared spectrum covering 1.2–2.3 μm is obtained by focusing a carrier-envelope phase (CEP) stabilized idler pulse (1.7 μm) from an optical parametric amplifier (Quantronix HE-TOPAS pumped by Spectra-Physics Spitfire-Ace) onto a hollow fibre set within a Kr-filled chamber (Femtolasers). Pulse compression is performed using both active mirror (OKO, 19-ch linear MMDM) and chirped mirror (Femtolasers) techniques. The pulse width is derived from the SHG autocorrelation.

**Transient reflectivity and transmittance measurements.**

We performed transient reflectivity and transmittance measurements for the single crystal and the thin film [27] using a 6 fs pulse. In this study, the intensity of the pump pulse is controlled in the wide range from 0.01 to 1.3 mJ/cm$^2$ with the step of 0.006 mJ/cm$^2$ by a pair of wire-grid CaF$_2$ polarizers[30]. In the transient reflectivity/transmittance measurement, the probe pulse reflected/transmitted from the sample is detected by an InGaAs detector (New-Focus model 2034) after passing through a spectrometer (JASCO, M10).

**Data Availability**

The data that support the plots within this paper and other findings of this study are available from the corresponding author upon reasonable request.

Acknowledgements

This work was supported by Japan Science and Technology Agency [Core Research for Evolutional Science and Technology (CREST: Elucidation of elementary dynamics of photoinduced phase transition by using advanced ultrashort light pulses) and Exploratory Research for Advanced Technology (ERATO: JPMJER1301)] and Japan Society for the Promotion of Science (JP15H02100, JP23244062, JP16K13814, JP17K14317, JP26887003, JP16K05459, JP25287080, JP26287070, JP17H02916, JP16H04140). Part of the work was conducted in the Equipment Development Center (Institute for Molecular Science), supported by the Nanotechnology Platform Program (Molecule and Material Synthesis) of Ministry of Education, Culture, Sports, Science and Technology, Japan.

Author's contributions

Y. K., T. A, Y. Y., Y. A, H. I., and S. Iwai developed the 6-fs light source and carried out the transient reflectivity/transmittance measurements using the 6-fs pulse and analysed the data with contribution from H. K.. G. K., and H. M. Y. performed the synthesis and the characterization of the thin film. K. I., H. K., and T. S. performed the synthesis and the characterization of the single crystal. S. Ishihara, Y. T., and K. Y. made theoretical considerations and calculations. S. Iwai devised all the experiments. Y. K., K. Y., and S. Iwai wrote the paper after discussing with all the co-authors.

## Corresponding author

Correspondence to Shinichiro Iwai

Figure Legends

Fig. 1 **Temperature-$t/U_{\text{dimer}}$ phase diagram and non-linear charge oscillation**

**a** Temperature-$t/U_{\text{dimer}}$ phase diagram of κ-(BEDT-TTF)$_2$X [20-23], which is extracted based on controlling the chemical pressure (Supplementary-1). $T_{\text{SC}}$ and $T_{\text{END}}$ respectively indicate the superconducting transition temperature and the critical end point of the first order Mott transition line. **b** Charge excitations from bonding to anti-bonding states are induced by weak light, while non-linear polar charge oscillation is driven by strong light-field. The purple clouds show charge distribution in the triangular dimer structure.

Fig. 2 **Transient reflectivity($\Delta R/R$) and transmittance ($\Delta T/T$) spectra**

**a** $\sigma$ spectra of single crystalline κ-(BEDT-TTF)$_2$Cu[N(CN)$_2$]Br (4 K). The orange line shows the analysis using the Drude-Lorentz model [25, 26] consisting of the dimer band (red line), the Hubbard band (blue line), Drude response (green line), and another band (magenta line). **b** Steady-state reflectivity (R) and transient reflectivity ($\Delta R/R$) spectra of single crystal with polarization along $c$-axis for $I_{\text{ex}} = 1.0$ mJ/cm$^2$ at $t_{\text{d}} = 10$ fs (red line), 200 fs (magenta line), and 500 fs (purple line). The $\Delta R/R$ spectrum at $t_{\text{d}} = 200$ fs is well reproduced by the Lorentz analysis (gray dashed line) assuming an additional oscillator (oscillation energy 0.62 eV, damping energy 0.04 eV, Supplementary-2). **c** $\Delta T/T$ (blue dots) and $\Delta R/R$ (red dots) spectra of 180 nm thin film κ-(BEDT-TTF)$_2$Cu[N(CN)$_2$]Br for $I_{\text{ex}} = 1.0$ mJ/cm$^2$ at $t_{\text{d}} = 10$ fs. $\Delta R/R>0$ and $\Delta T/T>0$ are ascribed to a stimulated emission (S. E.)

Fig. 3 **Non-linearity of transient reflectivity**

**a** ΔR/R spectra for $t_d$=10 fs at $I_{ex}$=1.0, 0.1 (×10), and 0.01 mJ/cm$^2$(×20), respectively, at 6 K. **b** Time profiles of ΔR/R peak at $I_{ex}$=1.0 mJ/cm$^2$ (red line) and 0.1 mJ/cm$^2$ (blue line) are reproduced using the conventional method as shown by the black line (Supplementary-5). **c** Oscillating component obtained by subtracting the fitting curve from the time profile in **b**. The black line shows a cosine oscillation (period :43 fs, damping: 70 fs, initial phase, -0.15π )(Supplementary-5). **d** $I_{ex}$ dependence of ΔR/R at 0.63 eV for $t_d$=10 fs and 500 fs. The dashed and solid lines show eye guides. $I_{ex}$ dependences of ΔR/R for larger $I_{ex}$ (additionally for $t_d$=200 fs, orange circles) are shown in the inset.

Fig. 4 **Anomalous enhancement of ΔR/R near $T_{SC}$ and $T_{END}$**

**a, b** Temperature dependence of ΔR/R spectrum for $t_d$=10 fs at $I_{ex}$ =1.0 mJ/cm$^2$ (**a**) and 0.01 mJ/cm$^2$ (**b**). **c, d, e** Temperature dependence of ΔR/R at 0.62 eV for $t_d$=10 fs, for $I_{ex}$ =1.0 mJ/cm$^2$ (**c**), 0.1 mJ/cm$^2$ (**d**) and 0.01 mJ/cm$^2$ (**e**). The dashed (white in a and b, red in c-e) lines show $T_{SC}$ and $T_{END}$, respectively.

Fig. 5 **Origin of non-linear charge oscillation in 2D extended Hubbard model (Supplementary-8 )**

**a** Schematic polar charge oscillation in κ-(BEDT-TTF)$_2$X with bonds $b_1$, $b_2$, $p$, $q$. **b** Directions of charge motions through different bonds indicated by arrows. **c** Upper panel: Calculated optical conductivity spectrum (polarization along $c$-axis). Lower panel: Calculated FT spectra of charge-density time profile of a molecule for $F$=0.01 (green dots), 0.10 (blue dots), and 0.16(red dots). The energy scale is slightly larger than the experimental one owing to the smallness of the system. Also for finite $V_{ij}$(orange dashed line for $F$=0.16), the spectral nature is essentially the same. **d** Ratio between FT peak intensities at 0.7 eV ($I_{0.7}$) and 0.93 eV ($I_{0.93}$) in the lower panel of Fig. 5 **c** with $V_{ij}$=0 as a function of $U$. The increase in $I_{0.93}/I_{0.7}$ for larger $U$ indicates that the non-linear charge oscillation becomes dominant.

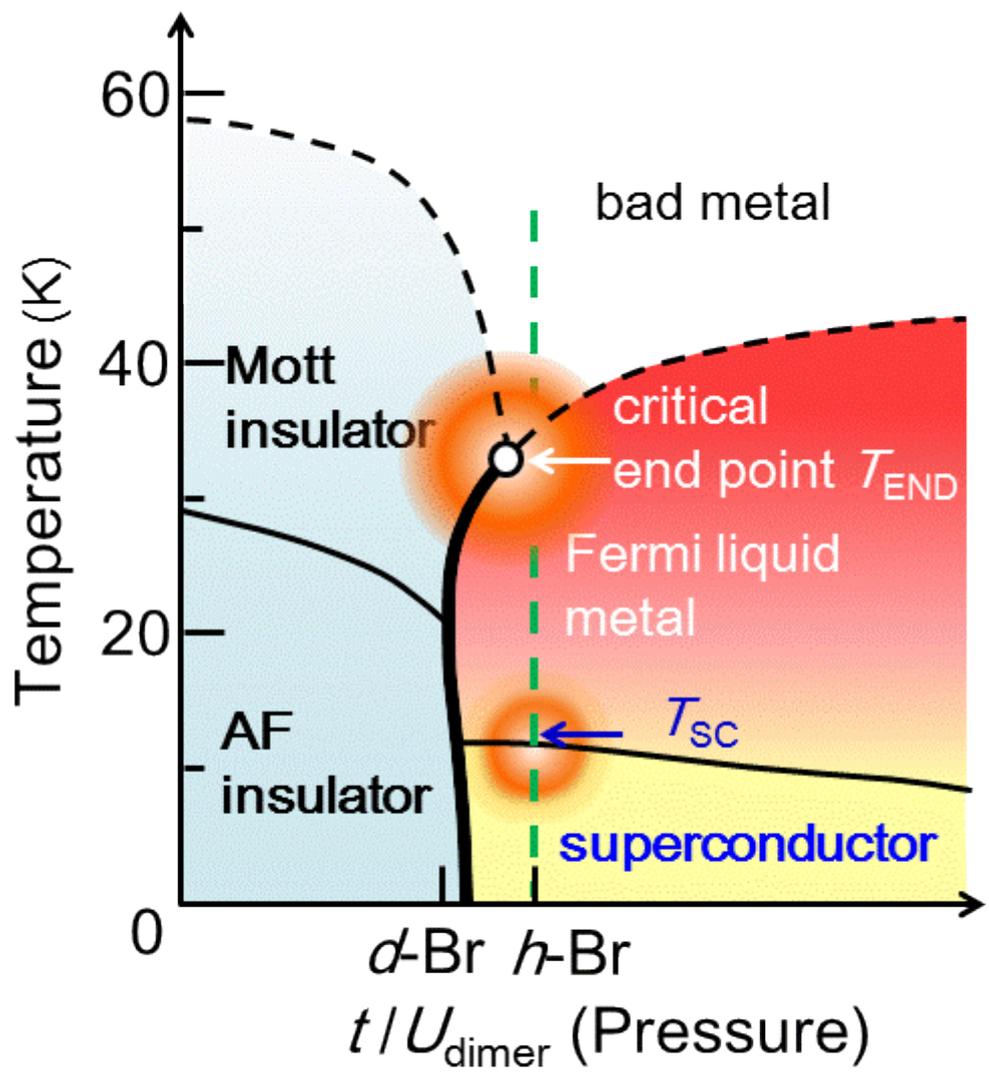

Kawakami et al. Fig. 1 a

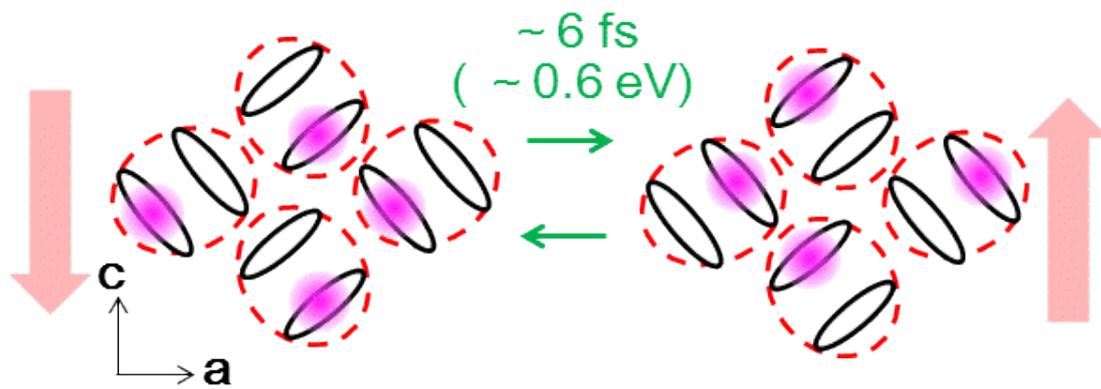

Non-linear charge oscillation (Strong field)

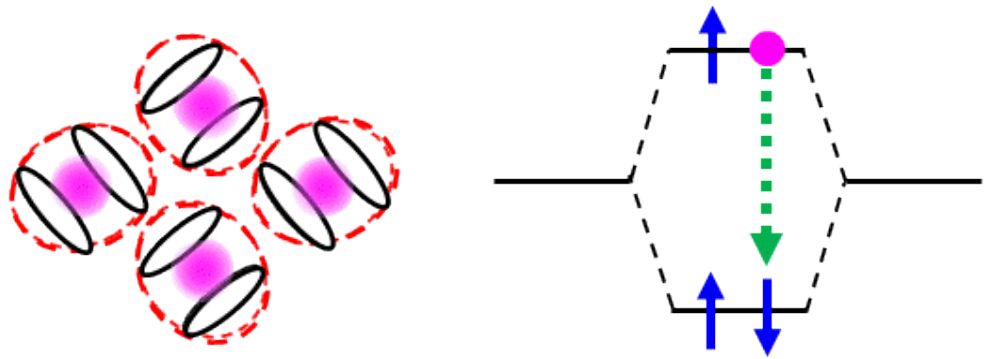

Bonding to anti-bonding transition (Weak field)

Kawakami et al. Fig. 1 b

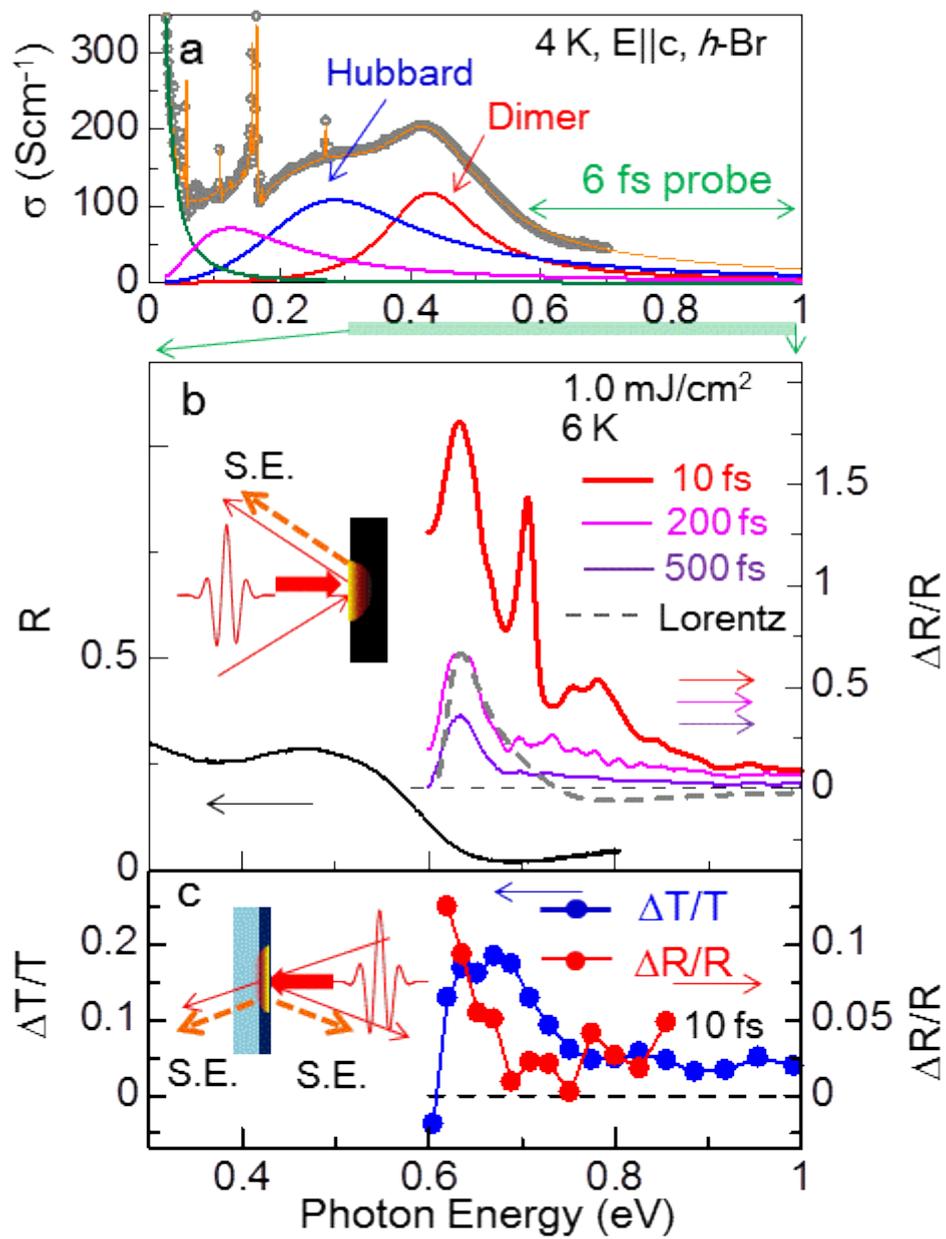

Kawakami et al. Fig. 2

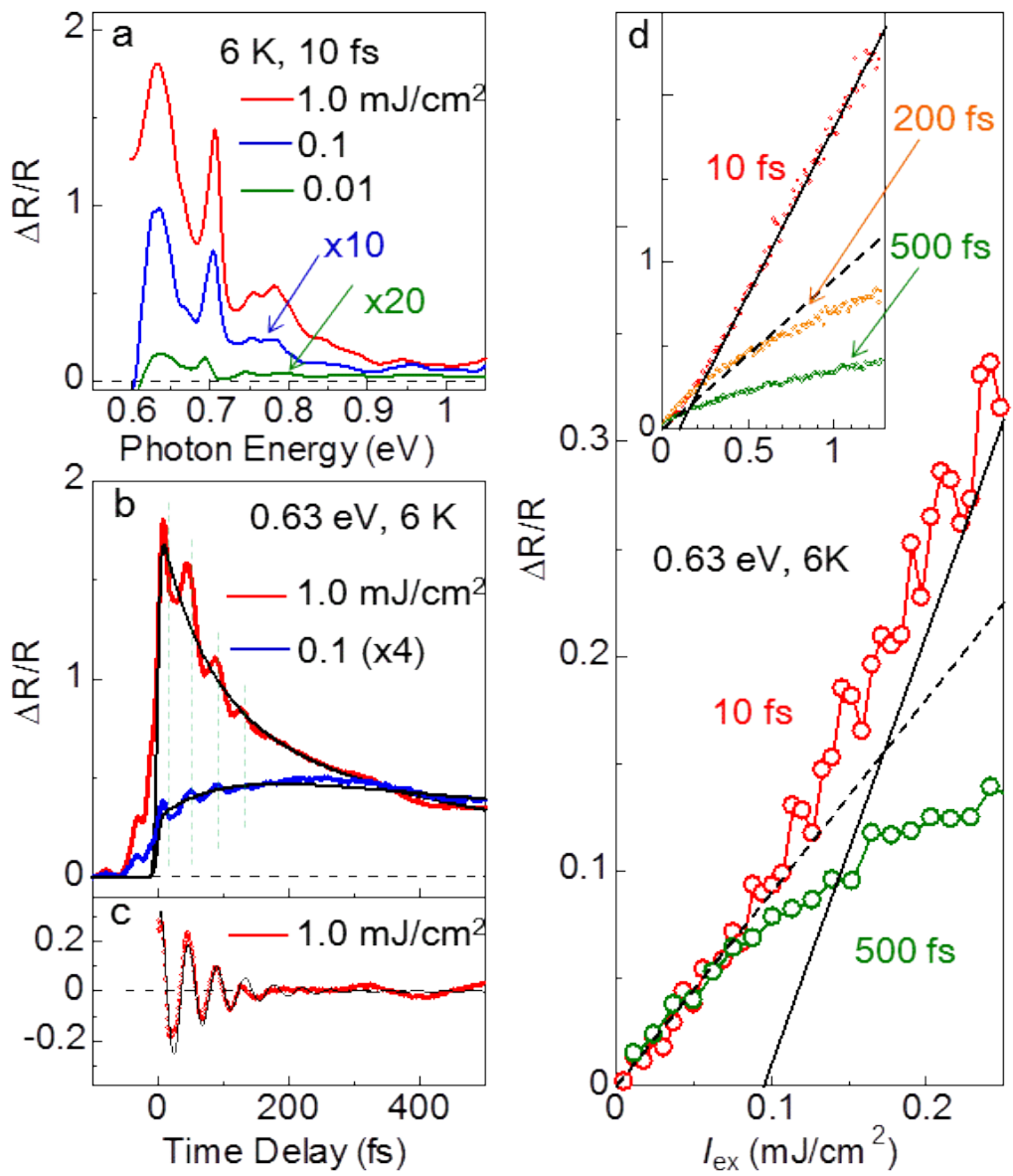

Kawakami et al.
Fig. 3

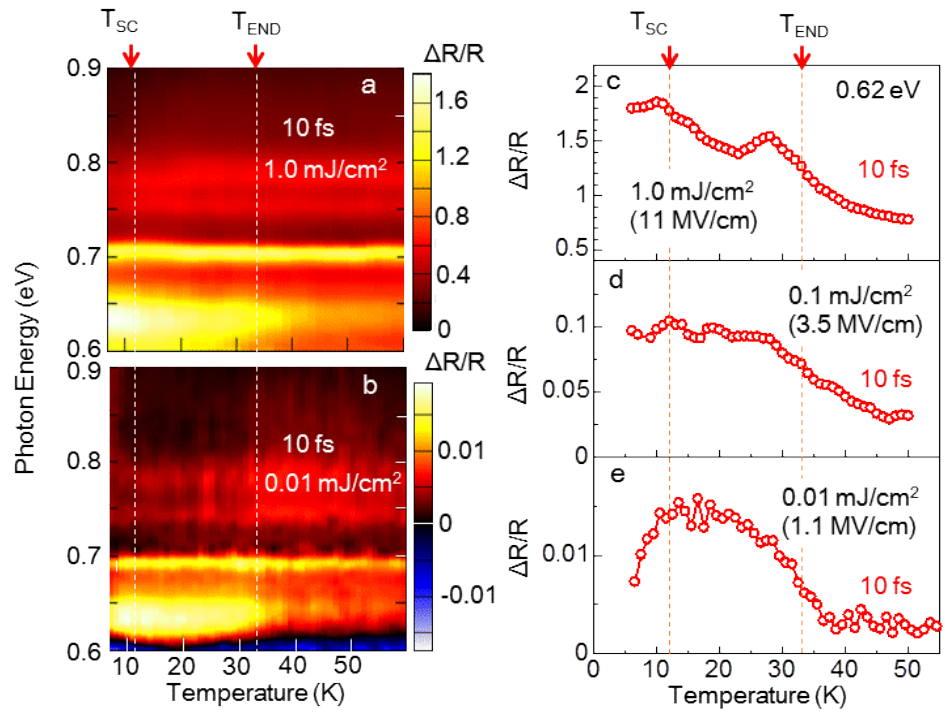

Kawakami et al.
Fig. 4

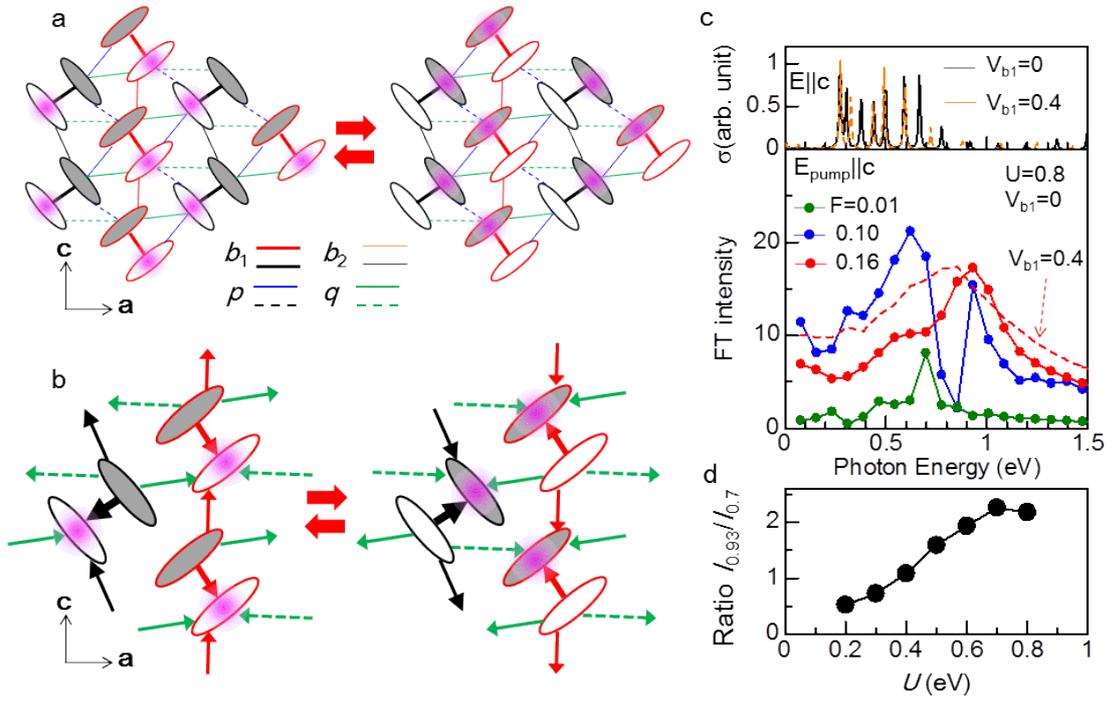

Kawakami et al. Fig. 5

Supplementary information for

# Nonlinear charge oscillation driven by a single-cycle light field in an organic superconductor

Y. Kawakami[1], T. Amano[1], Y. Yoneyama[1], Y. Akamine[1], H. Itoh[1],

G. Kawaguchi[2], H. M. Yamamoto[2], H. Kishida[3], K. Itoh[4], T. Sasaki[4],

S. Ishihara[1], Y. Tanaka[5], K. Yonemitsu[5] and S. Iwai[1*]

[1]Department of Physics, Tohoku University, Sendai, Japan
[2]Institute for Molecular Science, Okazaki, Japan
[3]Department of Applied Physics, Nagoya University, Nagoya, Japan
[4]Institute for Materials Research, Tohoku University, Sendai, Japan
[5]Department of Physics, Chuo University, Tokyo, Japan

*Corresponding author: s-iwai@tohoku.ac.jp

### 1. Critical end point of the first order Mott transition line

$T_{END}$~33 K in the phase diagram of Fig. 1a [20-23] is actually obtained in chemically pressured $\kappa$-(d-BEDT-TTF)$_2$Cu[N(CN)$_2$]Br (d-Br) and $\kappa$-[(h-BEDT-TTF)$_{(1-x)}$(d-BEDT-TTF)$_x$]$_2$Cu[N(CN)$_2$]Br [S1]. Although the first order transition line is terminated around $x$ = 0.5 (30-35 K) in the phase diagram, an extrapolated line for $x$ < 0.5 (to the higher pressure region) has been discussed in terms of the gradual change from a bad metal to a Fermi liquid metal (in the sense that the Drude component is observed), as characterized by cusps in the temperature dependences of conductivity [S2], magnetic susceptibility [S3] and a sample volume [S4]. The anomalous increase in $\Delta R/R$ near $T_{END}$ is observed on this extrapolated line. In h-Br, the electronic phase crosses over from the bad metal to the Fermi liquid metal near $T_{END}$ shown in Fig. 1. From this fact, we expect that the observed anomaly at about 30 K in h-Br is concerned with $T_{END}$.

2. Spectral shape of $\Delta R/R$ at $t_d$ = 10 fs and 200 fs

The $\Delta R/R$ spectrum at $t_d$ = 200 fs (Fig. 2b) is well reproduced by the Lorentz analysis (grey dashed line) assuming an additional oscillator (oscillation energy 0.62 eV, damping energy 0.04 eV). The spectral width of the peak at $t_d$ = 200 fs [0.04 eV = $\hbar$/(100 fs)], which is much narrower than that of the dimer band (170 meV), is roughly consistent with the time constant of the fast decay component of 70 fs. This fact indicates that the stimulated emission is induced by only one electronic mode which is selectively driven in a coherent manner reducing dissipation. At shorter $t_d$ = 10 fs, the $\Delta R/R$ spectrum has a broad tail reflecting the uncertainty relation between time and frequency. A coherent artifact is also detected as a small 0.7 eV peak at $t_d$ = 10 fs.

3. Increase in electron and lattice temperatures

The application of an 11 MV cm$^{-1}$ light field is expected to increase electron and lattice temperatures. Considering the coefficient of the linear temperature dependent term of the specific heat $\gamma$ = 22 mJ K$^{-2}$ mol$^{-1}$ [S5], the electron temperature is finally expected to be 520 K. However, the rise in the electron temperature is negligible at $t_d$ = 10 fs [when the anomalies are observed in the temperature dependence (Fig. 4)] because electrons are scattered only a few times before in this compound. In fact, an anomaly is detected in Fig. 4c near $T_{SC}$. On the other hand, the lattice temperature at a millisecond after the pulse irradiation (when the next laser pulse arrives in a 1-kHz operation) is a little higher than before, resulting in a heat accumulation and the fact that the anomalies at ~10 K and ~28 K are shifted from $T_{SC}$ and $T_{END}$ to the low-temperature side by 2~4 K.

4. Transient transmittance measurement in a thin film of $\kappa$-(BEDT-TTF)$_2$Cu[N(CN)$_2$]Br

We should be careful about that the phase boundary between the superconducting and the insulating phases is sensitive to the pressure, i.e., a thin film of $\kappa$-(BEDT-TTF)$_2$Cu[N(CN)$_2$]Br ($h$-Br) can be made insulating (possibly with small superconducting fluctuations) owing to the negative pressure effect from the substrate [27]. If the thin film is insulating at all temperatures as a consequence (actually this is the case), it is impossible to obtain information on influences from the metal-insulator and superconducting transitions by measuring $\Delta T/T$ and $\Delta R/R$ on the thin film of $h$-Br. The nonlinear charge oscillation is observed not only for superconducting $h$-Br (single crystal) (Fig. 2b), but also for insulating $h$-Br (thin film) (Fig. 2c) and for insulating $\kappa$-($d$-BEDT-TTF)$_2$Cu[N(CN)$_2$]Br. ($d$-Br, single crystal) (inset of Fig. S1a), although the peak energy for $d$-Br (~0.69 eV) is slightly higher than that for $h$-Br (0.63 eV). It is quite reasonable because this oscillation emerges from the dimerized molecular structure and the consequent transfer integrals, although details in low-energy electronic structures are different.

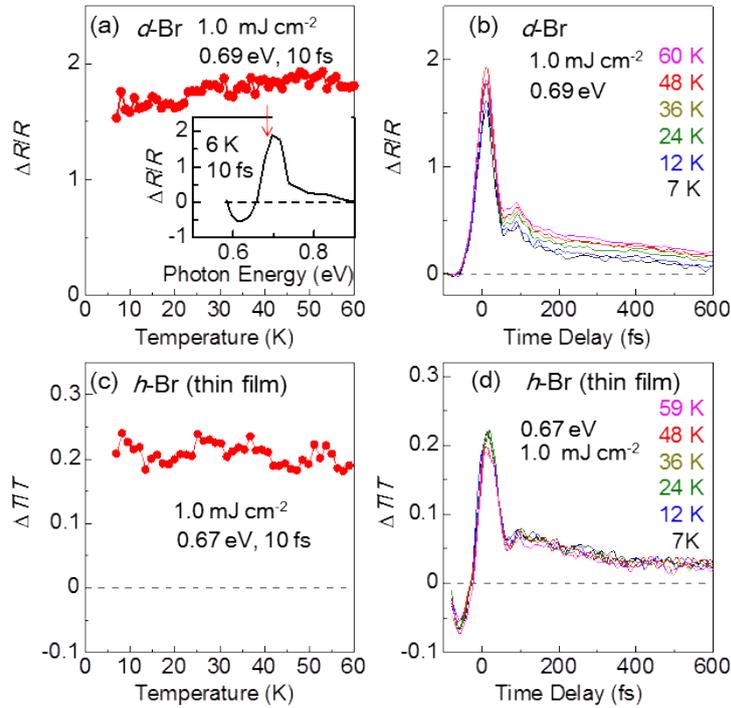

**Figure S1 Temperature dependence and time profiles of $\Delta R/R$ ($\kappa$-($d$-BEDT-TTF)$_2$Cu[N(CN)$_2$]Br) and of $\Delta T/T$ ($\kappa$-(BEDT-TTF)$_2$Cu[N(CN)$_2$]Br thin film). a** Temperature dependence of $\Delta R/R$ at 0.69 eV in a single crystal of $\kappa$-($d$-BEDT-TTF)$_2$Cu[N(CN)$_2$]Br ($d$-Br). The inset shows the $\Delta R/R$ spectrum for $t_d$ = 10 fs at 6 K. $I_{ex}$ = 1.0 mJ cm$^{-2}$. Excitation and detection polarizations are along the $c$-axis. **b** Time profiles of $\Delta R/R$ at 0.69 eV for $d$-Br. **c** Temperature dependence of $\Delta T/T$ at 0.67 eV in a thin film of $\kappa$-(BEDT-TTF)$_2$Cu[N(CN)$_2$]Br ($h$-Br). **d** Time profiles of $\Delta T/T$ at 0.67 eV in a thin film of $h$-Br.

Thus, we use the $h$-Br thin film for judging whether the reflectivity increase is due to a stimulated emission or an induced absorption. On the other hand, the anomalous enhancement of the nonlinear charge oscillation is observed only for superconducting $h$-Br (single crystal, Fig. 4), not detected for insulating $h$-Br (thin film, Fig. S1c) or $d$-Br (single crystal, Fig. S1a). No anomaly in the temperature dependence in $d$-Br (Fig. S1a), which does not have $T_{SC}$ or $T_{END}$ on its small $t/U_{dimer}$ side, shows that the distinctive enhancements of $\Delta R/R$ are related to $T_{SC}$ and $T_{END}$. The time profiles of $\Delta T/T$ for $h$-Br (thin film) (Fig. S1d) and those of $\Delta R/R$ for insulating $d$-Br (bulk) (Fig. S1b) are also shown.

## 5. Analysis of time profile of $\Delta R/R$

The time profile of the $\Delta R/R$ peak is analysed by the conventional method as shown by the black line (Fig. 3b) obeying the equation

$$\Delta R/R(t) = \int_{-\infty}^{\infty} K(t)G(t-t')dt'$$

$$K(t) = A_{fast}\left[1-\exp(-t/\tau_{f\text{-rise}})\right]\exp(-t/\tau_f) + \left[1-\exp(-t/\tau_{s\text{-rise}})\right]\sum_{n=1}^{3} A_{slow\,n}\exp(-t/\tau_{sn})$$

$$G(t) = \exp\left[-t^2/\left\{(4\ln 2/9)^2\right\}\right]$$

where the Gaussian $G(t)$ with a width of time resolution (= 9 fs) is an instrumental response. $A_{fast}$ and $A_{slow\,n}$ ($n = 1\sim3$) represent coefficients for the fast and slow components. For the red line in Fig.3b ($I_{ex}$ = 1.0 mJ cm$^{-2}$), the fast build-up ($\tau_{f\text{-rise}}$ < 5 fs) with decay ($\tau_f$ = 70 fs) is dominant (65.4%) in comparison with the slow rise ($\tau_{s\text{-rise}}$=90 fs) with three-component decay $\tau_{s1}$ = 360 fs (29.7%), $\tau_{s2}$ = 4.2 ps (1.3%), $\tau_{s3}$ > 100 ps (3.6%) and the oscillation (period 43 fs, damping time 70 fs) (Figs. 3b and 3c). For the blue line in Fig. 3b [$I_{ex}$ = 0.1 mJ cm$^{-2}$ (×4)], the fast component (with same $\tau_{f\text{-rise}}$ and $\tau_f$) is smaller (32.2%), while the slow rise is dominant (67.8%; $\tau_{s1}$ = 360 fs (20.8%), $\tau_{s2}$ = 2.2 ps (45.5%), $\tau_{s3}$ > 100 ps (1.5%)).

The oscillating component is shown in Fig. 3c, which is obtained by subtracting the fitting curve from the time profile in Fig. 3b. The black line shows a cosine oscillation (period: 43 fs, damping: 70 fs, initial phase: -0.15$\pi$). The 43 fs [$\sim\hbar/(0.1$ eV)] oscillation in Figs. 3b and 3c is attributed to a coherent intra-molecular vibration [$\nu_{60}(B_{3g})$] which strongly interacts with the electronic state [26, S6].

## 6. Temperature dependence of $\Delta R/R$ for $t_d = 500$ fs

The temperature dependence of $\Delta R/R$ at 0.62 eV for $t_d = 500$ fs is shown by the blue squares for $I_{ex}$ = 1.0, 0.1 and 0.01 mJ cm$^{-2}$ in Figs. S2a, S2b and S2c, respectively [The data for $t_d = 10$ fs (red circles) have already been shown in Fig. 4 of the main text]. The 500 fs response is also sensitive to superconducting fluctuations below ~$2T_{SC}$.

It is noted that the enhancement of $\Delta R/R$ near $T_{SC}$ is observed even at $t_d = 10$ fs (only for a high-excitation intensity) as shown in Fig. 4. However, as shown above, the 500 fs response is also sensitive to superconducting fluctuations below ~$2T_{SC}$ (even for weaker excitations $I_{ex}$ = 0.1 and 0.01 mJ cm$^{-2}$ in Figs. S2b and S2c). Such a picosecond response has already been discussed in terms of quasi-particle dynamics which is related to the pseudo gap in cuprates [10-12, 14] and this compound [13, 14].

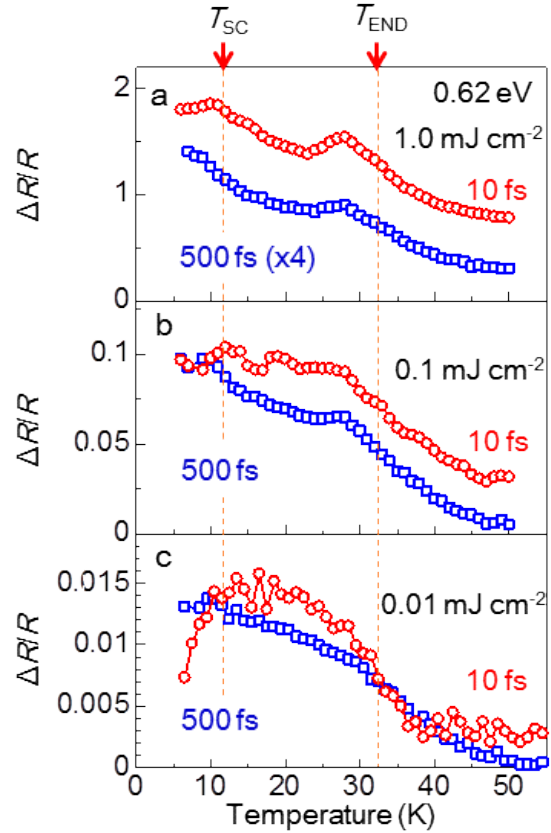

**Figure S2 Temperature dependences of $\Delta R/R$ for $t_d = 500$ fs**

Temperature dependences of $\Delta R/R$ at 0.62 eV [$t_d$ = 500 (blue squares) and 10 fs (red circles), the data for $t_d = 10$ have already been shown in Fig. 4 of the main text], for $I_{ex}$ = 1.0 mJ cm$^{-2}$ (a), 0.1 mJ cm$^{-2}$ (b) and 0.01 mJ cm$^{-2}$ (c).

In contrast, the <10 fs ultrafast rise (the rise time obtained by the fitting is <5 fs, as described in Supplementary 5) cannot be driven by such a quasi-particle dynamics that responds on the time scale of picosecond. In fact, $\Delta R/R$ indicates a very fast (70 fs) decay component, showing that the nonlinear charge oscillation is damped before the conventional quasi-particle dynamics plays a dominant role. Considering the ultrafast rise (<10 fs corresponding to the energy scale of >0.4 eV), the Coulomb repulsive interaction should play an important role. It is also demonstrated by the theoretical calculation that the nonlinear charge oscillation becomes dominant with increasing on-site Coulomb energy (Fig. 5d).

In addition, the anomaly in $\Delta R/R$ near $T_{SC}$ for $t_d$ = 10 fs shows the $I_{ex}$ dependence which is different from that for $t_d$ = 500 fs, i.e., the increase in $\Delta R/R$ with 10 fs-rise/70 fs-decay below ~2$T_{SC}$ is clear for $I_{ex}$ = 1 mJ cm$^{-2}$ (Fig. 4c), while it is substantially reduced for 0.1 mJ cm$^{-2}$ (Fig. 4d) and turns to a decrease for 0.01 mJ cm$^{-2}$ (Fig. 4e). In contrast, such a distinct $I_{ex}$-dependence of the anomaly cannot be detected for $t_d$ = 500 fs (the increase toward $T_{SC}$ is almost independent of $I_{ex}$ for $t_d$ = 500 fs). Such a difference in the $I_{ex}$ dependence also indicates that the origin of the response at $t_d$ = 10 fs is completely different from that at $t_d$ = 500 fs.

# 7. Temperature dependence of steady state $R$ in $\kappa$-(BEDT-TTF)$_2$Cu[N(CN)$_2$]Br

Anomalies near $T_{SC}$ and $T_{END}$ observed for both $t_d$ = 10 fs and 500 fs are not seen for the steady state reflectivity in this spectral region (Fig. S3).

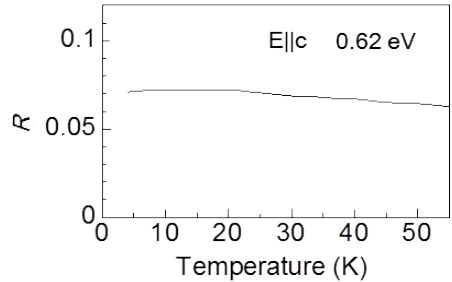

**Figure S3 Temperature dependence of $R$ in $\kappa$-(BEDT-TTF)$_2$Cu[N(CN)$_2$]Br**

Temperature dependence of steady state reflectivity at 0.62 eV.

## 8. Technical details of theoretical consideration (2D extended Hubbard model):

To clarify the origin of the nonlinear charge motion which is detected as a $\Delta R/R$ peak at 0.63 eV, we theoretically investigated charge oscillations which are driven by the single-cycle light field in a two-dimensional 3/4-filled extended Hubbard model

$$H_{2D} = \sum_{\langle i,j \rangle \sigma} t_{ij} \left( c_{i\sigma}^{+} c_{j\sigma} + c_{j\sigma}^{+} c_{i\sigma} \right) + U \sum_{i} n_{i\uparrow} n_{i\downarrow} + \sum_{\langle i,j \rangle} V_{ij} n_i n_j \qquad (1)$$

where $c_{i\sigma}(c_{i\sigma}^{+})$ is the annihilation (creation) operator of an electron on site $i$ with spin $\sigma$, $n_{i\sigma} = c_{i\sigma}^{+} c_{i\sigma}$, and $n_i = \sum_{\sigma} n_{i\sigma}$. This model has on-site Coulomb repulsion ($U$), nearest-neighbor repulsions ($V_{ij}$) and transfer integrals between sites $i$ and $j$ ($t_{ij}$) for $b_1$, $b_2$ $p$, $q$ bonds for a 16-site system with periodic boundary conditions as shown in Fig. 5a [S7]. The intermolecular distances and angles are taken from the structural data for $\kappa$-(BEDT-TTF)$_2$Cu[N(CN)$_2$]Br at 20 K [S8, S9]. We use $t_{b1}$ = -0.2850 eV, $t_{b2}$ = -0.1062 eV, $t_p$ = -0.1136 eV, and $t_q$ = 0.0400 eV, which are estimated from the extended Hückel calculation [S9], $U$ = 0.8 eV, $V_{b1}$ = 0.0 eV, $V_{b2}$ = 0.0 eV, $V_p$ = 0.0 eV, and $V_q$ = 0.0 eV for $F$ = 0.01, 0.10, and 0.16, and $U$ = 0.8 eV, $V_{b1}$ = 0.40 eV, $V_{b2}$ = 0.24 eV, $V_p$ = 0.28 eV, and $V_q$ = 0.24 eV for $F$ = 0.16 in Fig. 5c. Photoexcitation by the single-cycle light field with an amplitude $F$ [V/angstrom] along the $c$-axis is introduced through the Peierls phase [S10].

Charge oscillations are calculated with the help of the time-dependent Schrödinger equation starting from the exact ground state. The Fourier transform (FT) spectra of the time-dependent charge density of a certain molecule (the FT spectra are independent of the molecule) are obtained for the time span $T < t < 10T$ with $T = 2\pi / \omega_c$, where $\hbar \omega_c$ = 0.7 eV is the central frequency of the light field. The steady state optical conductivity ($\sigma$) for polarization along the $c$-axis is calculated as before [S11].